# IceCube – Astrophysics and Astroparticle Physics at the South Pole

THE ICECUBE COLLABORATION

## IceCube Collaboration Member List

R. Abbasi[28], Y. Abdou[22], T. Abu-Zayyad[33], M. Ackermann[39], J. Adams[16], J. A. Aguilar[28], M. Ahlers[32], M. M. Allen[36], D. Altmann[1], K. Andeen[28,a], J. Auffenberg[38], X. Bai[31,b], M. Baker[28], S. W. Barwick[24], R. Bay[7], J. L. Bazo Alba[39], K. Beattie[8], J. J. Beatty[18,19], S. Bechet[13], J. K. Becker[10], K.-H. Becker[38], M. L. Benabderrahmane[39], S. BenZvi[28], J. Berdermann[39], P. Berghaus[31], D. Berley[17], E. Bernardini[39], D. Bertrand[13], D. Z. Besson[26], D. Bindig[38], M. Bissok[1], E. Blaufuss[17], J. Blumenthal[1], D. J. Boersma[1], C. Bohm[34], D. Bose[14], S. Böser[11], O. Botner[37], A. M. Brown[16], S. Buitink[14], K. S. Caballero-Mora[36], M. Carson[22], D. Chirkin[28], B. Christy[17], F. Clevermann[20], S. Cohen[25], C. Colnard[23], D. F. Cowen[36,35], A. H. Cruz Silva[39], M. V. D'Agostino[7], M. Danninger[34], J. Daughhetee[5], J. C. Davis[18], C. De Clercq[14], T. Degner[11], L. Demirörs[25], F. Descamps[22], P. Desiati[28], G. de Vries-Uiterweerd[22], T. DeYoung[36], J. C. Díaz-Vélez[28], M. Dierckxsens[13], J. Dreyer[10], J. P. Dumm[28], M. Dunkman[36], J. Eisch[28], R. W. Ellsworth[17], O. Engdegård[37], S. Euler[1], P. A. Evenson[31], O. Fadiran[28], A. R. Fazely[6], A. Fedynitch[10], J. Feintzeig[28], T. Feusels[22], K. Filimonov[7], C. Finley[34], T. Fischer-Wasels[38], B. D. Fox[36], A. Franckowiak[11], R. Franke[39], T. K. Gaisser[31], J. Gallagher[27], L. Gerhardt[8,7], L. Gladstone[28], T. Glüsenkamp[39], A. Goldschmidt[8], J. A. Goodman[17], D. Góra[39], D. Grant[21], T. Griesel[29], A. Gross[16,23], S. Grullon[28], M. Gurtner[38], C. Ha[36], A. Haj Ismail[22], A. Hallgren[37], F. Halzen[28], K. Han[39], K. Hanson[13,28], D. Heinen[1], K. Helbing[38], R. Hellauer[17], S. Hickford[16], G. C. Hill[28], K. D. Hoffman[17], A. Homeier[11], K. Hoshina[28], W. Huelsnitz[17,c], J.-P. Hülss[1], P. O. Hulth[34], K. Hultqvist[34], S. Hussain[31], A. Ishihara[15], E. Jacobi[39], J. Jacobsen[28], G. S. Japaridze[4], H. Johansson[34], K.-H. Kampert[38], A. Kappes[9], T. Karg[38], A. Karle[28], P. Kenny[26], J. Kiryluk[8,7], F. Kislat[39], S. R. Klein[8,7], J.-H. Köhne[20], G. Kohnen[30], H. Kolanoski[9], L. Köpke[29], S. Kopper[38], D. J. Koskinen[36], M. Kowalski[11], T. Kowarik[29], M. Krasberg[28], T. Krings[1], G. Kroll[29], N. Kurahashi[28], T. Kuwabara[31], M. Labare[14], K. Laihem[1], H. Landsman[28], M. J. Larson[36], R. Lauer[39], J. Lünemann[29], J. Madsen[33], A. Marotta[13], R. Maruyama[28], K. Mase[15], H. S. Matis[8], K. Meagher[17], M. Merck[28], P. Mészáros[35,36], T. Meures[13], S. Miarecki[8,7], E. Middell[39], N. Milke[20], J. Miller[37], T. Montaruli[28,d], R. Morse[28], S. M. Movit[35], R. Nahnhauer[39], J. W. Nam[24], U. Naumann[38], D. R. Nygren[8], S. Odrowski[23], A. Olivas[17], M. Olivo[10], A. O'Murchadha[28], S. Panknin[11], L. Paul[1], C. Pérez de los Heros[37], J. Petrovic[13], A. Piegsa[29], D. Pieloth[20], R. Porrata[7], J. Posselt[38], P. B. Price[7], G. T. Przybylski[8], K. Rawlins[3], P. Redl[17], E. Resconi[23,e], W. Rhode[20], M. Ribordy[25], M. Richman[17], J. P. Rodrigues[28], F. Rothmaier[29], C. Rott[18], T. Ruhe[20], D. Rutledge[36], B. Ruzybayev[31], D. Ryckbosch[22], H.-G. Sander[29], M. Santander[28], S. Sarkar[32], K. Schatto[29], T. Schmidt[17], A. Schönwald[39], A. Schukraft[1], A. Schultes[38], O. Schulz[23,f], M. Schunck[1], D. Seckel[31], B. Semburg[38], S. H. Seo[34], Y. Sestayo[23], S. Seunarine[12], A. Silvestri[24], G. M. Spiczak[33], C. Spiering[39], M. Stamatikos[18,g], T. Stanev[31], T. Stezelberger[8], R. G. Stokstad[8], A. Stössl[39], E. A. Strahler[14], R. Ström[37], M. Stüer[11], G. W. Sullivan[17], Q. Swillens[13], H. Taavola[37], I. Taboada[5], A. Tamburro[33], A. Tepe[5], S. Ter-Antonyan[6], S. Tilav[31], P. A. Toale[2], S. Toscano[28], D. Tosi[39], N. van Eijndhoven[14], J. Vandenbroucke[7], A. Van Overloop[22], J. van Santen[28], M. Vehring[1], M. Voge[11], C. Walck[34], T. Waldenmaier[9], M. Wallraff[1], M. Walter[39], Ch. Weaver[28], C. Wendt[28], S. Westerhoff[28], N. Whitehorn[28], K. Wiebe[29], C. H. Wiebusch[1], D. R. Williams[2], R. Wischnewski[39], H. Wissing[17], M. Wolf[23], T. R. Wood[21], K. Woschnagg[7], C. Xu[31], D. L. Xu[2], X. W. Xu[6], J. P. Yanez[39], G. Yodh[24], S. Yoshida[15], P. Zarzhitsky[2], M. Zoll[34]

[1]*III. Physikalisches Institut, RWTH Aachen University, D-52056 Aachen, Germany*
[2]*Dept. of Physics and Astronomy, University of Alabama, Tuscaloosa, AL 35487, USA*
[3]*Dept. of Physics and Astronomy, University of Alaska Anchorage, 3211 Providence Dr., Anchorage, AK 99508, USA*
[4]*CTSPS, Clark-Atlanta University, Atlanta, GA 30314, USA*
[5]*School of Physics and Center for Relativistic Astrophysics, Georgia Institute of Technology, Atlanta, GA 30332, USA*
[6]*Dept. of Physics, Southern University, Baton Rouge, LA 70813, USA*
[7]*Dept. of Physics, University of California, Berkeley, CA 94720, USA*
[8]*Lawrence Berkeley National Laboratory, Berkeley, CA 94720, USA*
[9]*Institut für Physik, Humboldt-Universität zu Berlin, D-12489 Berlin, Germany*
[10]*Fakultät für Physik & Astronomie, Ruhr-Universität Bochum, D-44780 Bochum, Germany*
[11]*Physikalisches Institut, Universität Bonn, Nussallee 12, D-53115 Bonn, Germany*
[12]*Dept. of Physics, University of the West Indies, Cave Hill Campus, Bridgetown BB11000, Barbados*
[13]*Université Libre de Bruxelles, Science Faculty CP230, B-1050 Brussels, Belgium*
[14]*Vrije Universiteit Brussel, Dienst ELEM, B-1050 Brussels, Belgium*
[15]*Dept. of Physics, Chiba University, Chiba 263-8522, Japan*
[16]*Dept. of Physics and Astronomy, University of Canterbury, Private Bag 4800, Christchurch, New Zealand*
[17]*Dept. of Physics, University of Maryland, College Park, MD 20742, USA*
[18]*Dept. of Physics and Center for Cosmology and Astro-Particle Physics, Ohio State University, Columbus, OH 43210, USA*
[19]*Dept. of Astronomy, Ohio State University, Columbus, OH 43210, USA*
[20]*Dept. of Physics, TU Dortmund University, D-44221 Dortmund, Germany*
[21]*Dept. of Physics, University of Alberta, Edmonton, Alberta, Canada T6G 2G7*
[22]*Dept. of Physics and Astronomy, University of Gent, B-9000 Gent, Belgium*
[23]*Max-Planck-Institut für Kernphysik, D-69177 Heidelberg, Germany*
[24]*Dept. of Physics and Astronomy, University of California, Irvine, CA 92697, USA*
[25]*Laboratory for High Energy Physics, École Polytechnique Fédérale, CH-1015 Lausanne, Switzerland*
[26]*Dept. of Physics and Astronomy, University of Kansas, Lawrence, KS 66045, USA*
[27]*Dept. of Astronomy, University of Wisconsin, Madison, WI 53706, USA*
[28]*Dept. of Physics, University of Wisconsin, Madison, WI 53706, USA*
[29]*Institute of Physics, University of Mainz, Staudinger Weg 7, D-55099 Mainz, Germany*
[30]*Université de Mons, 7000 Mons, Belgium*
[31]*Bartol Research Institute and Department of Physics and Astronomy, University of Delaware, Newark, DE 19716, USA*
[32]*Dept. of Physics, University of Oxford, 1 Keble Road, Oxford OX1 3NP, UK*
[33]*Dept. of Physics, University of Wisconsin, River Falls, WI 54022, USA*
[34]*Oskar Klein Centre and Dept. of Physics, Stockholm University, SE-10691 Stockholm, Sweden*
[35]*Dept. of Astronomy and Astrophysics, Pennsylvania State University, University Park, PA 16802, USA*
[36]*Dept. of Physics, Pennsylvania State University, University Park, PA 16802, USA*
[37]*Dept. of Physics and Astronomy, Uppsala University, Box 516, S-75120 Uppsala, Sweden*
[38]*Dept. of Physics, University of Wuppertal, D-42119 Wuppertal, Germany*
[39]*DESY, D-15735 Zeuthen, Germany*
[a]*now at Dept. of Physics and Astronomy, Rutgers University, Piscataway, NJ 08854, USA*
[b]*now at Physics Department, South Dakota School of Mines and Technology, Rapid City, SD 57701, USA*
[c]*Los Alamos National Laboratory, Los Alamos, NM 87545, USA*
[d]*also Sezione INFN, Dipartimento di Fisica, I-70126, Bari, Italy*
[e]*now at T.U. Munich, 85748 Garching & Friedrich-Alexander Universität Erlangen-Nürnberg, 91058 Erlangen, Germany*
[f]*now at T.U. Munich, 85748 Garching, Germany*
[g]*NASA Goddard Space Flight Center, Greenbelt, MD 20771, USA*

**Acknowledgments**

We acknowledge the support from the following agencies: U.S. National Science Foundation-Office of Polar Programs, U.S. National Science Foundation-Physics Division, University of Wisconsin Alumni Research Foundation, the Grid Laboratory Of Wisconsin (GLOW) grid infrastructure at the University of Wisconsin - Madison, the Open Science Grid (OSG) grid infrastructure; U.S. Department of Energy, and National Energy Research Scientific Computing Center, the Louisiana Optical Network Initiative (LONI) grid computing resources; National Science and Engineering Research Council of Canada; Swedish Research Council, Swedish Polar Research Secretariat, Swedish National Infrastructure for Computing (SNIC), and Knut and Alice Wallenberg Foundation, Sweden; German Ministry for Education and Research (BMBF), Deutsche Forschungsgemeinschaft (DFG), Research Department of Plasmas with Complex Interactions (Bochum), Germany; Fund for Scientific Research (FNRS-FWO), FWO Odysseus programme, Flanders Institute to encourage scientific and technological research in industry (IWT), Belgian Federal Science Policy Office (Belspo); University of Oxford, United Kingdom; Marsden Fund, New Zealand; Japan Society for Promotion of Science (JSPS); the Swiss National Science Foundation (SNSF), Switzerland; D. Boersma acknowledges support by the EU Marie Curie IRG Program; A. Groß acknowledges support by the EU Marie Curie OIF Program; J. P. Rodrigues acknowledges support by the Capes Foundation, Ministry of Education of Brazil; A. Schukraft acknowledges the support by the German Telekom Foundation; N. Whitehorn acknowledges support by the NSF Graduate Research Fellowships Program.

# IceCube – Astrophysics and Astroparticle Physics at the South Pole

HERMANN KOLANOSKI[1] FOR THE ICECUBE COLLABORATION[2]
[1] *Institut für Physik, Humboldt-Universität zu Berlin, D-12489 Berlin, Germany*
[2] *See special section in these proceedings (ICRC 2011).*
*Hermann.Kolanoski@desy.de*

**Abstract:** The IceCube Neutrino Observatory at the South Pole has been completed in December 2010. In this paper we describe the final detector and report results on physics and performance using data taken at different stages of the yet incomplete detector. No signals for cosmic neutrinos from point sources and diffuse fluxes have been found. Prospects of these searches, including the setup of multi-messenger programs, are discussed. The limits on neutrinos from GRBs, being far below model predictions, require a reevaluation of GRB model assumptions. Various measurements of cosmic ray properties have been obtained from atmospheric muon and neutrino spectra and from air shower measurements; these results will have an important impact on model developments. IceCube observed an anisotropy of cosmic rays on multiple angular scales, for the first time in the Southern sky. The unique capabilities of IceCube for monitoring transient low energy events are briefly discussed. Finally an outlook to planned extensions is given which will improve the sensitivities both on the low and high energy side. The IceCube contributions to this conference (ICRC 2011) can be found in [1].

## 1 Introduction

The main component of the IceCube Neutrino Observatory at the geographic South Pole is a 1-km$^3$ detector instrumented with optical sensors in the clear ice of the polar glacier at depths between 1450 and 2450 m. The installation of IceCube with all its components was completed in December 2010. The main purpose of IceCube is the detection of high energy neutrinos from astrophysical sources via the Cherenkov light of charged particles generated in neutrino interactions in the ice or the rock below the ice.

The basic motivation for the construction of IceCube is to contribute to answering the fundamental, still unanswered question of the origin of cosmic rays. If cosmic rays are accelerated in astronomical objects, like Supernova Remnants (SNR), Active Galactic Nuclei (AGN) or Gamma Ray Bursts (GRB), one expects the accelerated particles to interact with the accelerator environment leading mainly to pion production. The principle of such a reaction of an accelerated hadron $N$ with an ambient hadron or photon is:

$$N+N', \gamma \to X + \left\{ \begin{array}{l} \pi^+ \to \mu^+ \nu_\mu \to e^+ \nu_\mu \bar{\nu}_\mu \nu_e \ \ (+\text{c.c.}) \\ \pi^0 \to \gamma\gamma \end{array} \right. \tag{1}$$

While neutral pions decay to gammas which can be detected by satellite gamma detectors up to several 100 GeV and by Cherenkov gamma ray telescopes in the TeV range, the charged pion decays or other weak decays such as kaon decays lead to neutrinos with a similar energy spectrum. If the pion production happens in or near the accelerator one expects to observe neutrino point sources. Interactions on the interstellar or intergalactic radiation background would lead to a diffuse flux of neutrinos. Also the summed flux of many faint sources could be seen as diffuse flux. The highest energies in the diffuse flux are expected to be in the EeV range stemming from interactions of the highest energy cosmic rays with the photons of the Cosmic Microwave Background (CMB). The observation of these neutrinos could confirm that the cosmic rays are limited at energies of about $10^{20}$ eV by the so-called "Greisen-Zatsepin-Kuzmin cut-off" (GZK cut-off). The importance of the observation of neutrinos from astrophysical sources to constrain theoretical models was stressed in various talks at this conference, for example [2].

In the lowest part of the IceCube detector a subvolume called DeepCore is more densely instrumented lowering the energy threshold from about 1 TeV in most of the detector to about 10 GeV. This addition to the original detector design extends appreciably the physics reach of the observatory to atmospheric neutrino oscillation phenomena, WIMP searches at lower masses and improves the sensitivity for the detection of transient events like supernovae and GRBs.

IceTop, the surface component of IceCube, is an air shower array covering an area of 1 km$^2$. With this detector airshowers from primary particles in the energy range from about 300 TeV to above 1 EeV can be measured. The detector is primarily designed to study the mass composition of primary cosmic rays in the energy range from about $10^{14}$ eV to $10^{18}$ eV by exploiting the correlation between the shower energy measured in IceTop and the energy deposited by muons in the deep ice, see [3].

Table 1: List of the years when a certain configuration of IceCube (IC), IceTop (IT) and DeepCore (DC) became operational. The DC strings are also included in the numbers for IC. In this paper we will use abbreviations like IC40, IT40 for the constellation in 2008, for example.

| Year | IC strings | IT stations | DC strings |
|---|---|---|---|
| 2006 | 9 | 9 | - |
| 2007 | 22 | 26 | - |
| 2008 | 40 | 40 | - |
| 2009 | 59 | 59 | - |
| 2010 | 79 | 73 | 6+7 |
| 2011 | 86 | 81 | 8+12 |

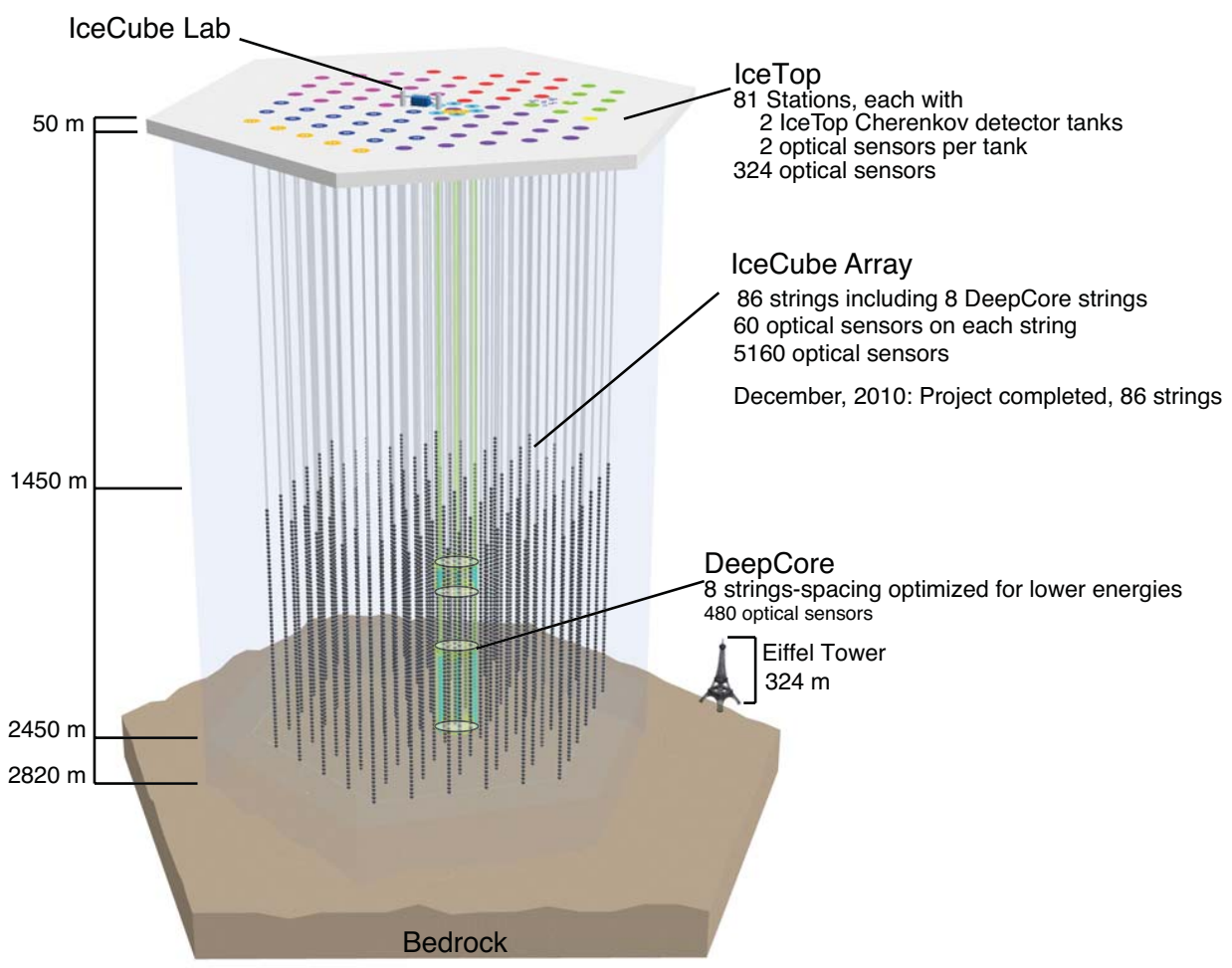

Figure 1: The IceCube detector with its components DeepCore and IceTop in the final configuration (January 2011).

In the following I will describe the IceCube detector with the sub-components DeepCore and IceTop. During the construction time from 2004 to the end of 2010 data have been taken with the still incomplete detector, see Table 1. Results obtained with differently sized detectors will be reported for neutrino point source searches, for diffuse neutrino fluxes searches, search for "exotic" (beyond standard model) particles and studies of cosmic rays. The summary includes a brief outlook to possible extensions in the future.

## 2 Detector

**IceCube:** The main component of the IceCube Observatory is an array of 86 strings equipped with 5160 light detectors in a volume of 1 km$^3$ at a depth between 1450 m and 2450 m (Fig. 1). The nominal IceCube string spacing is 125 m on a hexagonal grid (see DeepCore below).

Each standard string is equipped with 60 light detectors, called 'Digital Optical Modules' (DOMs), each containing a $10''$ photo multiplier tube (PMT) to record the Cherenkov light of charged particles traversing the ice. In addition, a DOM houses complex electronic circuitry supplying signal digitisation, readout, triggering, calibration, data transfer and various control functions [4]. The most important feature of the DOM electronics is the recording of the analog waveforms in $3.3\,\mathrm{ns}$ wide bins for a duration of $422\,\mathrm{ns}$. The recording is initiated if a pulse crosses a threshold of 0.25 photoelectrons. With a coarser binning a 'fast ADC' extends the time range to 6.4 $\mu$s.

**Ice properties:** At the depth of the detector the ice is very clear with an absorption length reaching up to 200 m. Scattering and absorption show a depth dependence, which follows the dust concentration in the polar glacier. The most prominent feature is a significantly higher dust concentration at depths between 2000 and 2100 m. The measurement and modelling of the ice properties for reconstruction and simulation is discussed in [5].

**DeepCore:** In the lower part of the detector a section called DeepCore is more densely instrumented. The DeepCore subarray includes 8 (6) densely instrumented strings optimized for low energies plus the 12 (7) adjacent standard strings (the numbers in parentheses apply to the DeepCore configuration of the 2010 running with 79 strings for which we will discuss results below).

**IceTop:** The 1-km$^2$ IceTop air shower array [3] is located above IceCube at a height of 2835 m above sea level, corresponding to an atmospheric depth of about 680 g/cm$^2$. It consists of 162 ice Cherenkov tanks, placed at 81 stations mostly near the IceCube strings (Fig. 1). In the center of the array, three stations have been installed at intermediate positions. Together with the neighbouring stations they form an in-fill array for denser shower sampling yielding a lower energy threshold. Each station comprises two cylindrical tanks, 10 m apart.

Each tank is equipped with two DOMs to record the Cherenkov light of charged particles that penetrate the tank. DOMs, electronics and readout scheme are the same as for the in-ice detector. The two DOMs in each tank are operated at different PMT gains to cover linearly a dynamic range of about $10^5$ with a sensitivity to a single photoelectron). The measured charges are expressed in units of 'vertical equivalent muons' (VEM) determined by calibrating each DOM with muons (see ref. [6]).

**Trigger and data acquisition:** The average noise rate of a single DOM is 540 Hz. To initiate the full readout of DOMs, a so-called 'hard local coincidence' (HLC) is required. In IceCube at least one of the two nearest neigbour DOMs of a string must have signals above threshold within $\pm 1\,\mu$s, resulting in a launch rate of 5-20 Hz per DOM, falling with increasing depth. In IceTop the HLC requirement is a coincidence of the two high gain DOMs of a sta-tion. This results in a launch rate of high gain DOMs of 2-4 Hz compared to about 1600 Hz of a single high gain DOM at a threshold of about 0.2 VEM.

In the counting house at the surface, triggers are formed from the HLCs deciding if the data are written to a permanent storage medium to make it available for later analysis. The basic in-ice trigger, for example, requires that at least 8 DOMs are launched by an HLC leading to a rate of about 2 kHz. A very loose trigger requirement is applied to the DOMs in the DeepCore fiducial region (below the dust layer) by requiring 3 or more HLC hits within a 2.5 $\mu$s time window. The basic trigger for IceTop is issued if the readouts of 6 or more DOMs are launched by an HLC leading to a rate of 30 to 40 Hz. For all detector components HLC hits are always stored in case of a trigger issued by another detector component.

For each DOM above threshold, even without a local coincidence, condensed data, so-called SLC hits ('soft local coincidence'), are transmitted. These data contain only the time and amplitude of the PMT waveform peaks (for in-ice DOMs) or the time and integrated charge (for IceTop DOMs). The SLC hits are, for example, used for detecting transient events and to generate vetos for special event signatures. In the case of IceTop they are useful for detecting single muons in showers where the electromagnetic component has been absorbed (low energies, outer region of showers, inclined showers).

For monitoring transient events via rate variations, the time of single hits are histogrammed. In IceTop the single hits in different tanks are obtained with various thresholds ('scaler rates' for heliosperic physics).

Triggered events which fulfill further criteria for various event classes ('muon', 'cascade' etc.) are sent via satellite to the IceCube Computing Center in Madison. In addition fast online processing produces alerts for other telescopes in case of significant neutrino accumulations (see Section 4.5 on follow-up programs).

## 3 Detection Methods and Performance

**IceCube performance:** Neutrino point source searches rely on directional information and are therefore performed with muon neutrinos. The directions of muon tracks can be reconstructed to 1° or better (see Moon shadow analysis [7]). Primarily IceCube is designed to measure up-going neutrinos[1] using the Earth as filter against the large background of high energy muons from cosmic rays. However, because the neutrino cross section increases with energy the Earth becomes opaque for neutrinos above about 1 PeV. This can be seen in Fig. 2 where the energy dependence of the effective area of IceCube is plotted for different zenith angles. The effective area is defined as the target area which yields the observed muon neutrino rate when each neutrino is detected with 100% probability. Since at high energies the background from down-going cosmic ray muons becomes relatively small IceCube has extended its search also to the Southern sky at high energies (see Section 4).

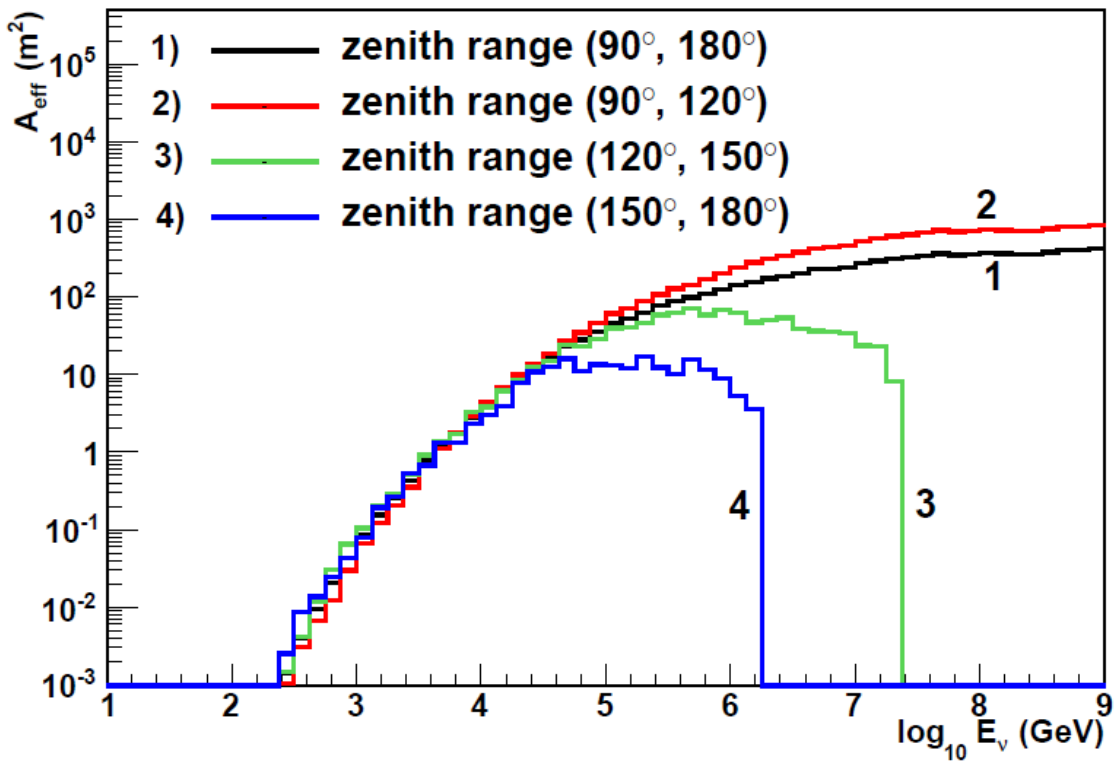

Figure 2: Effective area for muon neutrino detection with IceCube (IC40) as a function of the neutrino energy for different zenith angle ranges of the Northern sky. The plot was made for the IC40 analysis of diffuse neutrino fluxes [8].

For up-going neutrinos the background is mainly atmospheric neutrinos generated in the Northern atmosphere, while the background for down-going neutrinos comes mostly from high energy atmospheric muons. The extraction of signals for astrophysical neutrinos relies either on accumulations in space (point sources or extended objects) and/or in time (flares, GRB etc.) or on the assumption that the cosmic neutrino spectra are harder than for secondary cosmic rays, often a spectral index of about -2 compared to -3.7 for atmospheric muons and neutrinos is assumed. The latter is particularly important for the measurement of diffuse neutrino fluxes where an on/off source determination of background is not possible.

The muon neutrino energy cannot be directly measured (except in the cases where a neutrino interacts in the detector and the muon ranges out). The measured energy loss of muons in the ice is used as a proxy for the neutrino energy. For muon energies above about 1 TeV, corresponding to the critical energy of muons in ice, the total energy loss rises approximately linearly with the muon energy due to bremsstrahlung, pair production and nuclear interactions. This allows the determination of the muon energy from the energy loss with a resolution of about 50% ($\Delta \log_{10} E \approx 0.2$). The muon energy yields only a very coarse proxy for the neutrino energy which is only partially transferred to the muon. The angular resolution for muon neutrinos is about 1° at 1 TeV and about 0.5° at 1 PeV.

All flavours of neutrinos can contribute to the search of diffuse fluxes if they generate an electromagnetic or hadronic cascade in the ice. Electron and tau neutrinos can generate electromagnetic cascades in charged current interactions, all flavours can generate hadronic cascades via neutral current and charged current interactions. Cascades appear in IceCube as nearly spherical isotropic light sources, so that little direction information can be obtained. On the other

1. In this paper the term 'neutrinos' also includes anti-neutrinos.

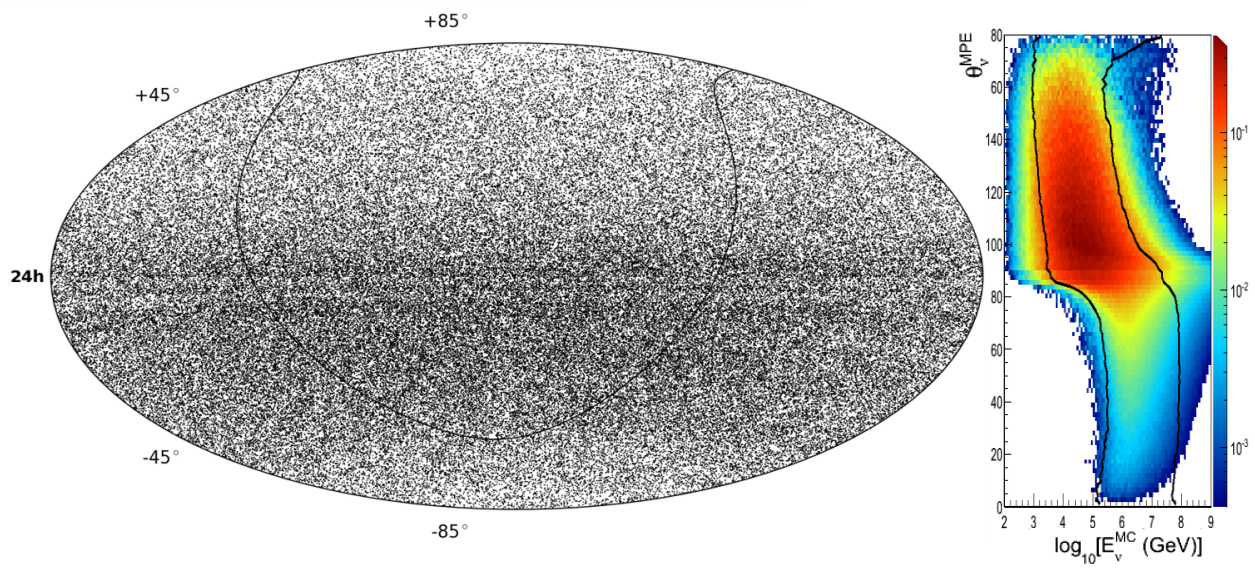

Figure 3: Skymap of neutrino candidates in equatorial coordinates for the IC40+59 data sample (left). The curved line indicates the galactic plane. The declination dependence of the selected energy ranges is shown on the right.

hand, the neutrino energy resolution is much better than in the case of neutrino detection via muons, about 30% at 10 PeV ($\Delta \log_{10} E \approx 0.13$). Neutrinos have to interact in or near the detector to be detectable as cascades which makes the neutrino effective area about an order of magnitude lower than for muon tracks.

**DeepCore performance:** The main improvement added by DeepCore is the decrease of the energy threshold to about 10 GeV. Using the surrounding IC strings as veto, one can identify low energy neutrino interactions inside or near the DC volume. In this way, atmospheric neutrinos can be collected over an angular range of $4\pi$ sr, yielding unprecedented statistical samples of atmospheric neutrinos, about 150,000 triggered atmospheric muon neutrinos per year, thus allowing oscillation studies [9]. The observation of a sizeable number of cascade event in DeepCore [10] confirms the performance expectations.

**IceTop performance:** IceTop will cover a primary energy range from about 300 TeV to 3 EeV for zenith angles up to about 65°. In coincidence with IceCube the zenith angle range is more limited yielding an angular coverage of about 0.3 sr. The event rate is sufficient for a composition analysis up to about 1 EeV.

The following resolutions have been obtained for 10 PeV (100 PeV) and for zenith angles smaller than 30° [11]: core position 7 m (8 m), zenith angle 0.5° (0.3°), energy resolution 0.05 (0.04) for $\log_{10} E/\mathrm{PeV}$.

## 4 Neutrino Point Sources

### 4.1 Search Strategy

The neutrino point source search relies on the precise directional information from muons generated by muon neutrinos interacting in the ice in and around the detector or the Earth crust below the detector. Figure 3 shows a skymap of arrival directions of neutrino candidates. The plot contains 57460 up-going and 87009 down-going neutrino candidates selected from 723 days of data taken with the 40 and 59 string configurations during 2008 and 2009 (IC40+59). In the Northern sky high energies are limited by neutrino absorption in the Earth (Fig. 2), in the Southern sky the energy threshold has to be increased to reject the large background from atmospheric muons.

In an unbiased search each direction has to be scanned leading to a large number of trials and thus a significance reduction. To improve signal significances one wants to reduce the number of trials by using additional information on the signal probabilities:

- Predefine a list of candidate sources which are theoretically likely to emit neutrinos.
- The list search can be further improved by summing the fitted signals for many sources ('stacking').
- Search for extended sources on scales from a few degrees, just resolvable, to scales of the size of the galactic plane.
- Search for time and spatial correlation with transient events, like flares in AGN.
- A special class of transient events are GRBs with particularly short emission times. The similar properties of different GRBs make them well suited for stacking (Section 4.4).
- Since IceCube is sensitive about 99% of the time to the full sky, alerts can be given to other telescopes if IceCube detects multiplets of neutrino candidates which accumulate in space and time. Such 'Follow-up Programs' are realised with optical, X-ray and $\gamma$-ray telescopes.

### 4.2 Full Sky Time-Integrated Search

In a basic approach one searches in the full considered data set for a significant accumulation of events in an angular range compatible with the angular resolution. For that purpose a likelihood function is defined which takes into account a possible signal and background:

$$L(n_s, \gamma) = \prod_{i=1}^{N} \left[ \frac{n_s}{N} S_i + \left(1 - \frac{n_s}{N}\right) B_i \right] \qquad (2)$$

For a given direction on the sky $S_i$ and $B_i$ are the probabilities for event $i$ to be signal or background, respectively; $N$ is the number of events which is looped over and $n_s$ is the number of most likely signal events. The likelihood function depends also on the energy via a spectral index $\gamma$ which is estimated in the search procedure. The search has to be done in a fine grid of directions, here at about 100 000 points, which reduces appreciably the "pre-trial" significance for a point source to a "post-trial" significance. The significances are evaluated defining a test statistics which compares the most likely values $\hat{n}_s, \hat{\gamma}$ with the null hypothesis. Using simulations the distribution of the test statistics for the case of no signal is evaluated yielding a p-value which is the probability to reach the observed or a higher significance for a result $\hat{n}_s, \hat{\gamma}$ if there is no signal.

In the analysis of the IC40+59 data the hottest spot at $(\mathrm{Ra, Dec}) = (75.45°, -18.15°)$ has a pre-trial p-value of $p_{pre} = 10^{-4.65}$, corresponding to an about 4 sigma significance, but a post-trial p-value of 0.67, indicating a high compatibility with the null hypothesis. This means that no significant point source observation can be reported from this search.

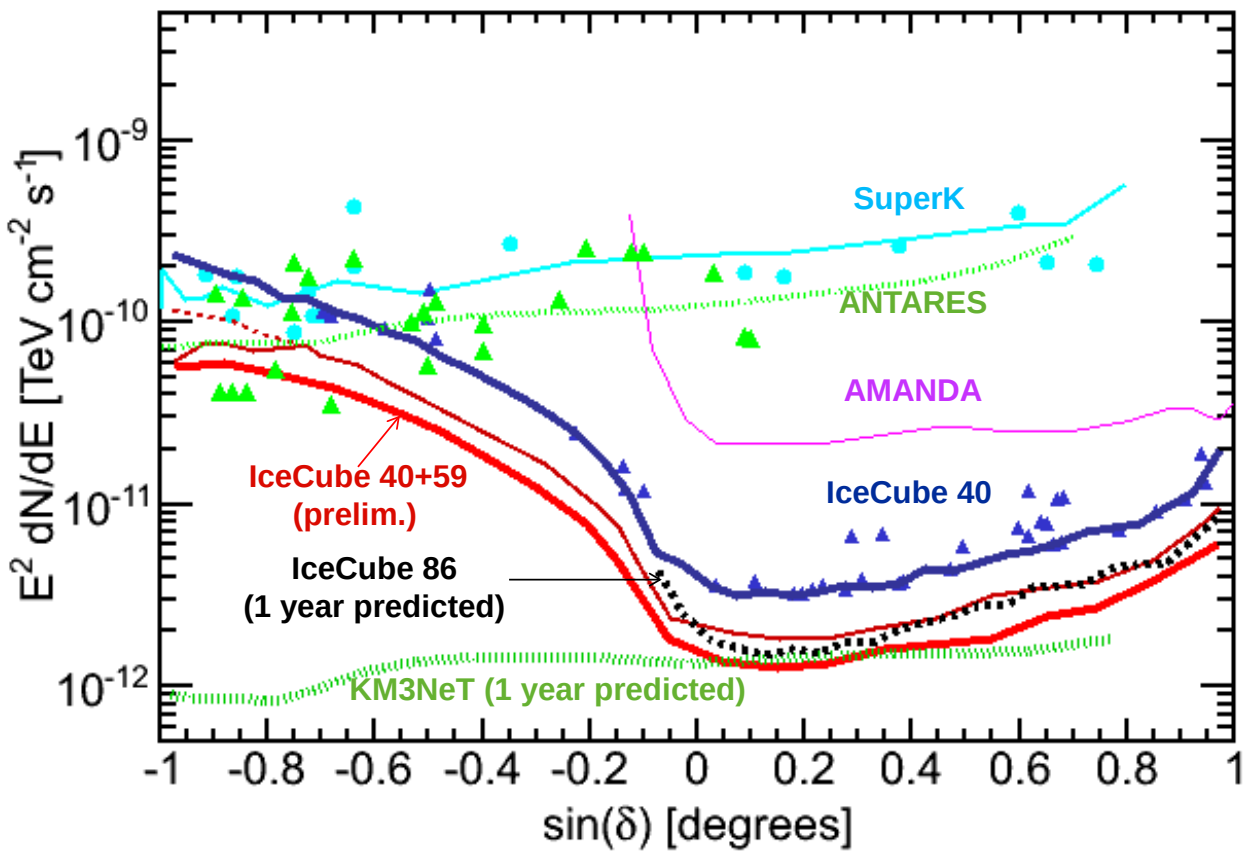

Figure 4: Neutrino point source limits (90% c.l.) for an $E^{-2}$ spectrum. The currently most stringent limit from IC40+59 data is compared to previous and expected limits.

An overview of limits obtained from time integrated point source searches is given in Fig. 4. The IceCube 40+59 results are compared to previously published limits from IceCube and other experiments. The recently published IC40 results [12] include also limits for specific source candidates which had been selected before looking at the data (see the list and more details in [12]). With these IceCube measurements the limits decreased by about a factor 1000 over the last 15 years.

The IC40+59 limits reached already the projected sensitivities obtainable by the full detector in one year. However, sensitivities below about $10^{-12}E^{-1}\mathrm{TeV}^{-1}\mathrm{cm}^{-2}\mathrm{s}^{-1}$ are necessary to seriously scrutinize models for cosmic ray acceleration with neutrino production. Hence in this search mode several years of additional data taking might be necessary for either a neutrino source discovery or a rejection of the models.

## 4.3 Time-Dependent Searches for Point Sources

The statistical significance can be improved by including time dependence in the likelihood function (2) since sources such as AGN can exhibit significant time variability in photon fluxes, which might also be visible in neutrinos, while the atmospheric background is roughly constant.

An example for an 'untriggered' search, i.e. without a priori time information, was presented at this conference [13] using the IC40+59 data as for the time-independent search (the IC40 analysis has recently been published [14]). The time-dependent likelihood term for this search is a Gaussian function, with its mean and width as free parameters. Scanning for flares of all durations from 20 seconds to 150 days the likelihood maximization returns the most significant flare from a particular direction. The strongest deviation from background was found in the IC59 data in a direction $(\mathrm{Ra}, \mathrm{Dec}) = (21.35°, -0.25°)$, centered on March 4, 2010 with a FWHM of 13 days (Fig. 5). An excess of 14 events is seen with a soft spectrum of $E^{-3.9}$, i.e. with no discrimination against the atmospheric spectrum. The post-trial p-value is determined to be 1.4% (corresponding to about 2.3 sigma) which is not sufficiently significant for claiming a neutrino flare discovery.

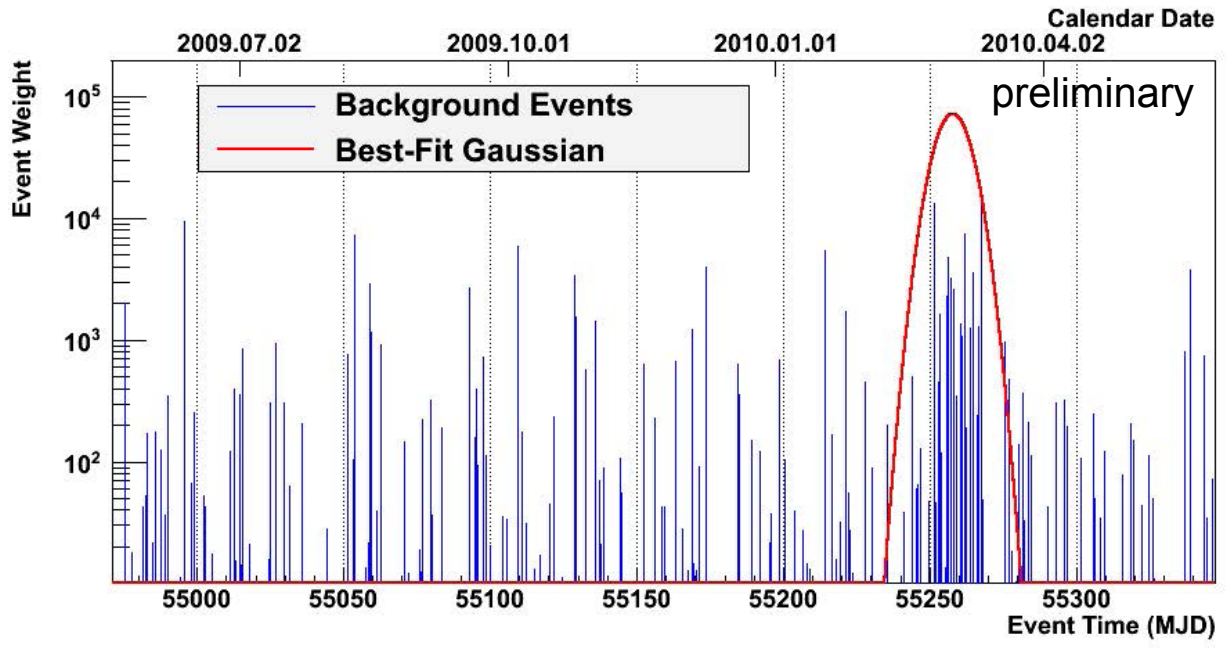

Figure 5: IC59: The time distribution of the signal-to-background ratio of events from the location of maximum significance. The curve is the fitted gaussian for the most significant flare.

## 4.4 Gamma Ray Bursts

The large energy dissipation in a Gamma Ray Burst (GRB) of about $10^{44}$ J suggests that a large fraction of the extragalactic cosmic rays at the highest energies could be accelerated in GRBs. GRBs are usually modeled as explosions of very massive stars which eventually collapse to a black hole. In such models the observed gamma rays stem from synchrotron radiation and/or inverse Compton scattering of electrons accelerated in shock fronts in the collimated explosive outflow. It was proposed that in the same way also protons are accelerated [15, 16]. These protons would undergo interactions with the surrounding photon field in the fireball and thus generate neutrinos according to (1). With their preferred parameters the models predict that GRB neutrinos be detectable by IceCube in less than one year.

At this conference a search using IC59 data was reported [17]. The search was based on a list of 98 GRB observations reported from satellites during times when IceCube was taking data. The neutrino search was done similarly to the point source searches, using a likelihood like (2) with an additional term for the time included. The time probability density function was flat in the interval were the first and last gamma rays were observed falling off smoothly to both sides. In the point-spread function the uncertainty in the GRB coordinates as obtained from satellites was included. No neutrino candidate was observed in the space-time windows. The analysis sets a limit far below the predicted model flux (Fig. 6). Combining the results from IC40 [18] and IC59 the corresponding upper limit lies a factor 5 be-low the model curve. This leads to the conclusion that either the model picture of GRBs is wrong or the chosen parameter values are not correct. Important model parameters are the Lorentz boost factor $\Gamma$ of the collimated outflow of the exploding star and the typical time scale $t_{var}$ of subsequent collisons of internal shocks. In [17] the limits obtained for the combination of these parameters are presented.

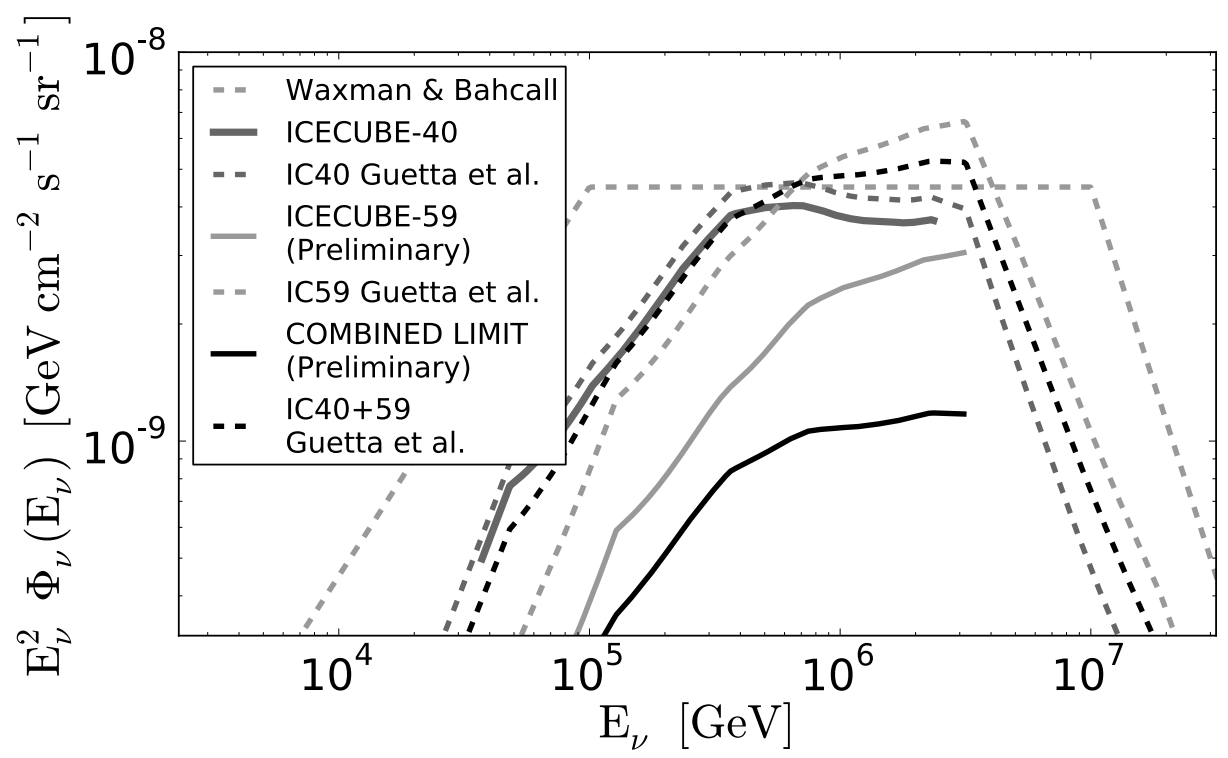

Figure 6: Limits on neutrino flux from GRBs compared to models from Waxmann [15] and Guetta et al. [16]. The derivation of the limits is based on the Guetta et al. model and accounts for the estimated properties of individual GRBs (the Waxmann predictions use average properties).

### 4.5 Follow-Up Programs

A special feature of the IceCube detector is that it is able to monitor the whole sky (though with different energy sensitivities, see Section 3). This can be exploited to send alerts to other telescopes with narrow fields of view (optical, X-ray, gamma-ray) if in a certain space-time window an excess of neutrinos above background is observed with a predefined significance. The alerted telescopes can then make follow-up observations on these 'targets of opportunity' which would lead to a significance enhancement if a positive correlation between different messenger signals are observed. The alert decisions have to be made fast, i.e. online at Pole and reported via satellite, and have to be tuned in a way that the alert rate is tolerable for the alerted partners.

The IceCube collaboration has follow-up programs established with several telescopes:

- Search for GRB and core-collapse supernovae: neutrino multiplets in a short time window, $< 100\,$s, generate alerts for optical follow-up by the Robotic Optical Transient Search Experiment (ROTSE) and the Palomar Transient Factory (PTF), see [19]. Furthermore an X-ray follow-up by the Swift satellite of the most significant multiplets has been set up and started operations in February 2011 [20].
- Search for neutrinos from TeV-gamma flares: a follow-up program with the MAGIC telescope has been tested with the IC79 setup and should become active for the IC86 running [21].

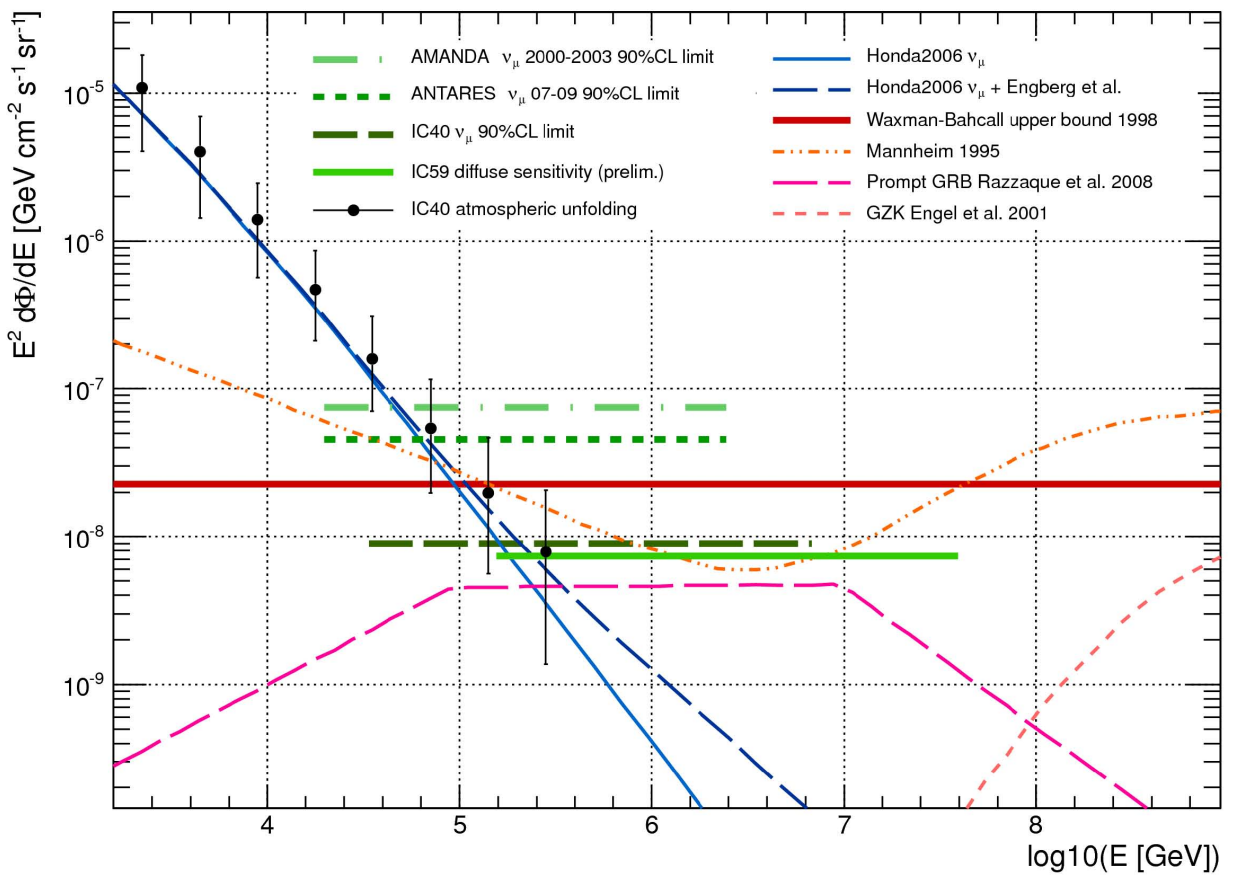

Figure 7: Limits and predictions for diffuse muon neutrino fluxes.

## 5 Diffuse Flux of Neutrinos

If there are many point sources, each with an unobservably low flux, then the aggregate flux may still be observable as a diffuse flux. Interactions of the cosmic rays with the matter and radiation near the source or somewhere else on their path through the space would lead, according to eq. (1), to meson production and the subsequent weak decays to a diffuse flux of neutrinos.

The identification of diffuse cosmic neutrinos relies on the assumption that they have a harder spectrum, e.g. $E^{-2}$ compared to about $E^{-3.7}$ for atmospheric neutrinos. The 'prompt' component of atmospheric neutrinos from decays of heavy flavour hadrons, which are produced predominantly in the first interactions in the atmosphere, is predicted to be harder than the 'conventional' neutrino flux. This introduces some additional uncertainty in the transition region where the cosmic flux is expected to become dominant. The experimental limits tell us that this transition is well above 100 TeV neutrino energy (see Fig. 7).

### 5.1 Diffuse Muon Neutrino Flux

Figure 7 shows the currently best limit obtained from IC40 data of up-going muons [22]. The points are the atmospheric neutrino spectrum determined by unfolding the measured muon energy depositions to obtain the flux as a function of the neutrino energy. The limit is now below the Waxmann-Bahcall bound [24] which gives a guideline of how much flux can be at most expected if cosmic neutrinos are generated in or near the accelerating sources (AGN, GRB etc.) via meson production as in (1).

### 5.2 Cascades and All-Flavour Neutrino Flux

The interaction of electron neutrinos in IceCube generates an electromagnetic cascade which manifests itself in the detector as a nearly spherical source of light with little information about the direction. To this 'cascade chan-nel' also neutral current interactions of all flavours, generating hadronic cascades, contribute. Therefore the results for cosmic neutrinos are expressed as all-flavour neutrino fluxes assuming a flavour ratio of 1:1:1 at the detector (evolving from a 1:2:0 ratio at the source by mixing). For diffuse flux measurements, the inferior directional resolution is not a major draw-back, but the relatively good energy resolution has substantial advantages (see Section 3).

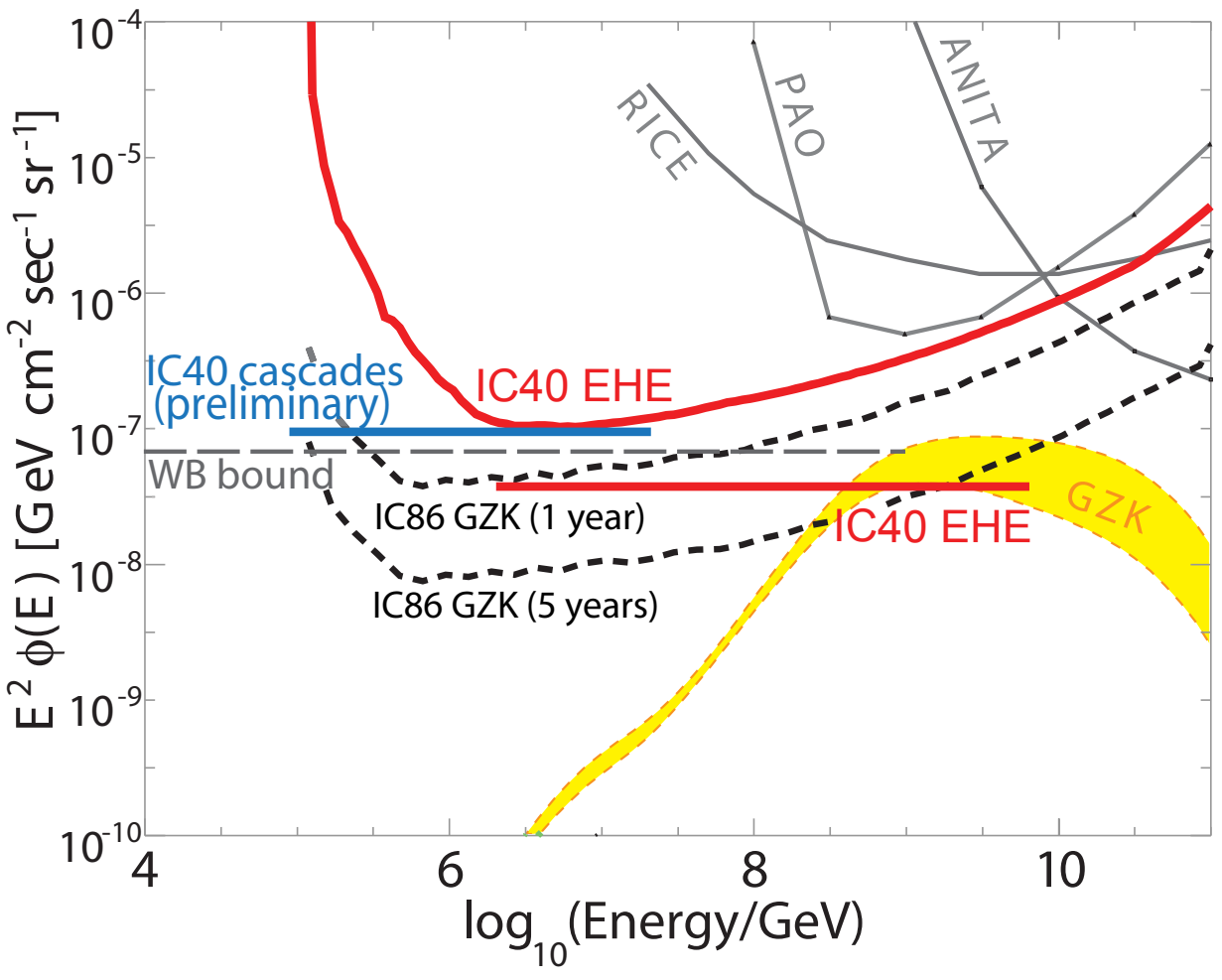

Figure 8: All-flavor flux limit from IC40 data derived from a cascade analysis [25] (blue line) and from an EHE neutrino search [27, 28] (red lines). The model dependent limits assuming an $E^{-2}$ neutrino spectrum are shown as horizontal lines. The curved red line is the model independent quasi-differential limit (normalized by decade in energy). The EHE neutrino flux limit is shown together with limits of other experiments (employing radio techniques). The estimated reach for the full detector in 1 and 5 years (black dashed lines) is compared to the specific model [23] shown as band on the plot.

At this conference a result on a cascade analysis using IC40 data taken over 374 days was presented [25]. Above a cascade energy cut of 25 TeV, 14 events are left in data while atmospheric neutrino simulations predict less than 4 events. Visual inspections of the experimental events indicate that at least a fraction of the excess events are induced by atmospheric muons. Because of insufficient statistics in the atmospheric muon background simulations and because of possible inaccuracies of the atmospheric neutrino flux prediction, this result is, at this moment, inconclusive (see also [26]).

The current best limit from cascades as derived in a more conservative analysis of the IC40 data for an energy range between about 90 TeV and 20 PeV is shown in Fig. 8.

### 5.3 Extremely-High Energy Neutrinos

Interactions of the highest energy cosmic rays with the photons of the Cosmic Microwave Background (CMB) are predicted to generate a diffuse neutrino flux in the EeV range. The observation of these neutrinos could confirm that the cosmic rays are limited at energies of about $10^{20}$ eV by the "GZK cut-off", at the point where the $\gamma_{\mathrm{CMB}}$ - nucleon system surpasses the threshold for pion production (with a strong enhancement due to the $\Delta$-resonance close to threshold). Since all involved processes and particles are well known this GZK process could be considered a 'guaranteed' source of cosmogenic neutrinos. However, in detail the theoretical predictions for the fluxes vary by about 3 orders of magnitude, depending mostly on the assumed primary composition and the distribution of cosmic ray sources.

At this conference results for 'Extremely-High Energy' (EHE) neutrinos using the IC40 detector have been presented [27, 28]. The analysis aims at finding down-going neutrinos generating very bright events in the detector. However, the large atmospheric muon background restricts the search to events coming from near the horizon where neutrinos have also the largest interaction probability. The obtained EHE neutrino flux limits are shown in Fig. 8. Up to about 10 EeV IceCube has the best upper limits. The comparison with predictions shows that a positive observation of GZK neutrinos might still take some years. On the other hand improvements in the analysis procedure could increase the detection efficiency [29]. For example a scheme is currently investigated to use single-tank hits in IceTop for a veto against the overwhelming background from downgoing muons [30].

## 6 Exotics

An essential part of the IceCube physics program deals with generic Particle Physics problems such as the search for new particles beyond the Standard Model, called 'exotic particles'. Such particles include Dark Matter candidates such as proposed by Supersymmetry (SUSY) or by Kaluza-Klein models. The breaking of larger symmetries, as postulated by 'Grand Unified Theories', implies the generation of topological defects such as monopoles which can also be searched for with IceCube.

### 6.1 WIMP Search

It is now experimentally well established that 'Dark Matter' (DM) exceeds normal, baryonic matter by about a factor 5. In most common scenarios the DM consists of Weakly Interacting Massive Particles (WIMPs) which remained from the Big Bang after the expansion rate of the Universe surpassed their annihilation rate. A promising WIMP candidate is the lightest supersymmetric particle, in most SUSY variants the neutralino $\chi$. In the searches reported below parameters within the MSSM ('Minimal Supersymmetric Model') have been investigated.

There are three general DM search strategies: In direct searches one looks for elastic WIMP scattering off nuclei; in indirect searches one tries to detect WIMP annnihilation products, such as gammas or neutrinos, with astropar-ticle detectors and finally in accelerator experiments one searches for pair-production of DM candidates. None of these searches has yet resulted in a conclusive discovery.

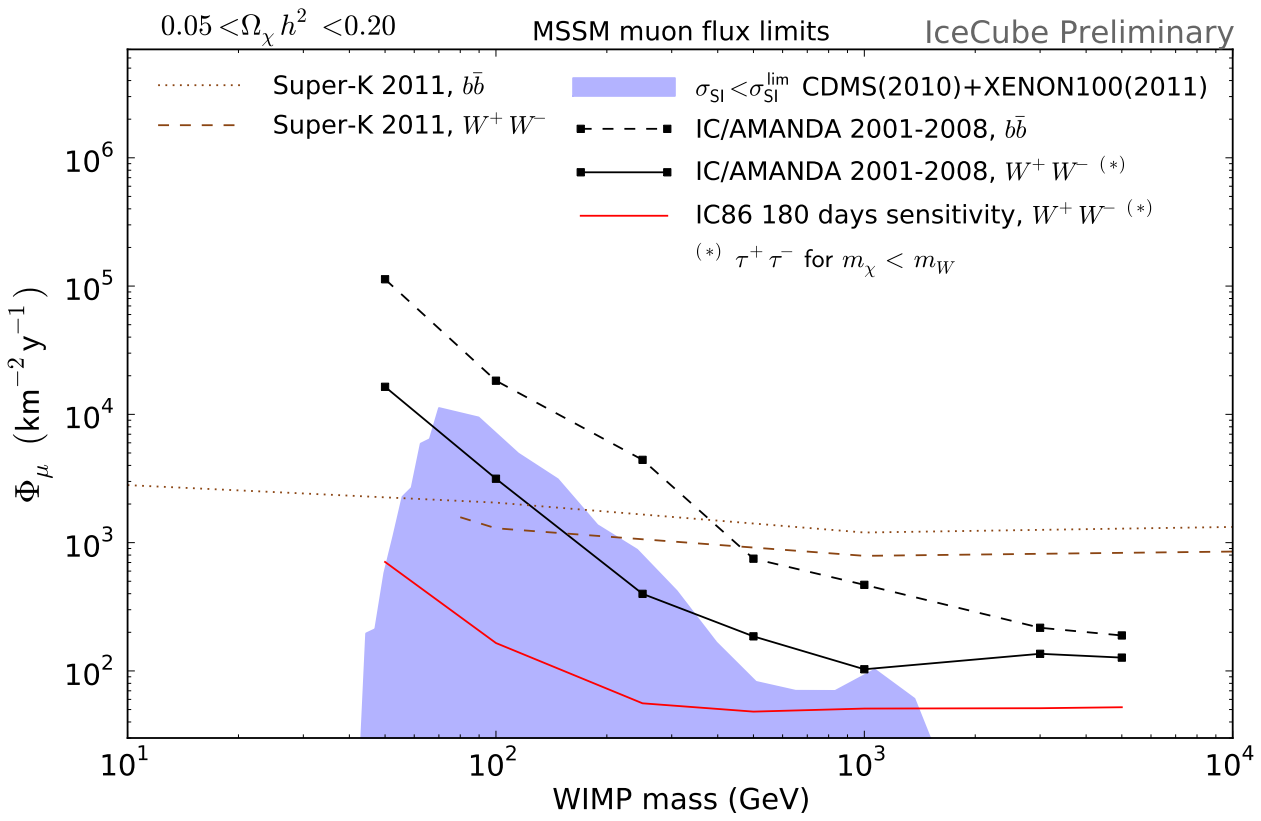

Figure 9: Preliminary limits on the WIMP induced muon flux from the Sun. See text for more explanations.

**WIMPs from the Sun:** IceCube is looking for neutrinos from WIMP (or other exotics) annihilation. The assumption is that WIMPs would accumulate in gravitational wells like the Sun or the center of the Milky Way. In Fig. 9 limits on an excess flux of muons from the Sun are given for WIMP masses from 50 GeV to 5 TeV [31]. The excess determination assumes a muon neutrino spectrum which depends on the WIMP mass and the annihilation channel. The studied channels $W^+W^-$ and $b\bar{b}$ have particularly hard and soft spectra, respectively (the harder the spectrum, the higher IceCube's sensitivity). The analysis combines data taken with IceCube and the precursor detector AMANDA between 2001 and 2008, with a total livetime of 1065 days when the Sun was below the horizon [31]. In 2008 AMANDA was taking data together with IceCube in the IC40 configuration employing an integrated data acquisition system. The figure also shows the estimated sensitivity for the full detector.

The muon flux limits can be related to direct measurements using the following chain of arguments: The accumulation requires a cool-down of the WIMPs by elastic scatters to get gravitationally trapped. The capture rate for WIMPs with a fixed mass can be calculated for given elastic cross sections (for different nuclei, with and without spin dependence) and the local WIMP density and velocity distribution. In this analysis the 'canonical' values $\rho_{\rm DM} = 0.3\,{\rm GeV/cm^3}$ and $\bar{v}_{\rm DM} = 270\,{\rm km/s}$ have been asummed. Assuming further that an equilibrium between capture and annihilation is reached in the Sun, the annihilation rate becomes half of the capture rate per WIMP. With further assumptions about the annihilation channels the rate of muons from muon neutrinos can be determined as described above. The shaded area in Fig. 9 indicates the region not yet excluded by the MSSM parameter constraints through the direct searches by the experiments CDMS and XENON100. For more details see [31].

**WIMP annihilation in the Milky Way and dwarf spheroidals:** In the contribution [32] a search for a neutrino excess from the galactic center and halo was reported. Using IC40 data (367 days) preliminary limits on $\langle \sigma_{ann}\, v \rangle$ have been obtained in the range $10^{-22}$ to $10^{-23}\,{\rm cm^3\,s^{-1}}$, depending on the WIMP mass. The 'natural scale', derived from cosmological parameters, is about $3 \cdot 10^{-26}\,{\rm cm^3\,s^{-1}}$. The limits depend strongly on the assumed model for the WIMP density and on the annihilation channel. A comparison of the limits for the $\tau^+\tau^-$ channel with the regions preferred by the satellite experiments PAMELA and Fermi (see details in [32]) shows that the WIMP searches of IceCube are constraining these preferred regions.

Another WIMP search reported at this conference is aiming at spheroidal dwarf galaxies [33] yielding the sensitivities for IC59 data. Although the sensitivities do not yet reach those of the $\gamma$-ray measurements (MAGIC, Fermi) the neutrino channel certainly adds complementary information.

### 6.2 Magnetic Monopoles

Relativistic magnetic monopoles, if they exceed the Cherenkov threshold at $\beta \approx 0.76$, deposit huge amounts of light in the detector and thus have a very clear signature. In a contribution to the conference a preliminary upper limit for the monopole flux was reported using IC22 data [34]. This limit is orders of magnitude better than previous ones.

Also discussed in [34] are the prospects of improving this limit with IC40 data to a level which is about a factor 1000 below the Parker bound. The Parker bound relates the observed strengths of cosmic magnetic fields to the maximal possible abundance of monopoles exploiting that the acceleration of monopoles by magnetic fields would damp those fields. The non-observation of monopoles constrains the combinations of Grand Unified Theories and inflationary scenarios.

## 7 Cosmic Rays

Origin, composition and spectrum of high energy cosmic rays are still not well understood. In particular above some 100 TeV, up to which direct measurements with balloons and satellitites are posssible, the experimental situation is far from being satisfactory. The understanding of the muon and neutrino fluxes from cosmic ray initiated air showers is also essential for IceCube because they are the major background in the search for extraterrestial neutrinos and exotic particles.

The IceCube observatory offers a variety of possibilities to measure cosmic rays, analyse composition and determine the spectra which can be used to tune the models. IceCube can be regarded as a cubic-kilometer scale three-dimensional cosmic ray detector with the air showers (mainly the electromagnetic component) measured by the surface detector IceTop and the high-energetic muons and neutrinos measured in the ice. In particular the mea-surement of the dominantly electromagnetic component of the airshower in IceTop in coincidence with the high energy muon bundle (muon threshold about 500 GeV), originating from the first interactions in the atmosphere, has a strong sensitivity to composition. Here IceCube offers the unique possibility to clarify the cosmic ray composition and spectrum in the range between about 300 TeV and 1 EeV, including the 'knee' region and a possible transition from galactic to extra-galactic origin of cosmic rays.

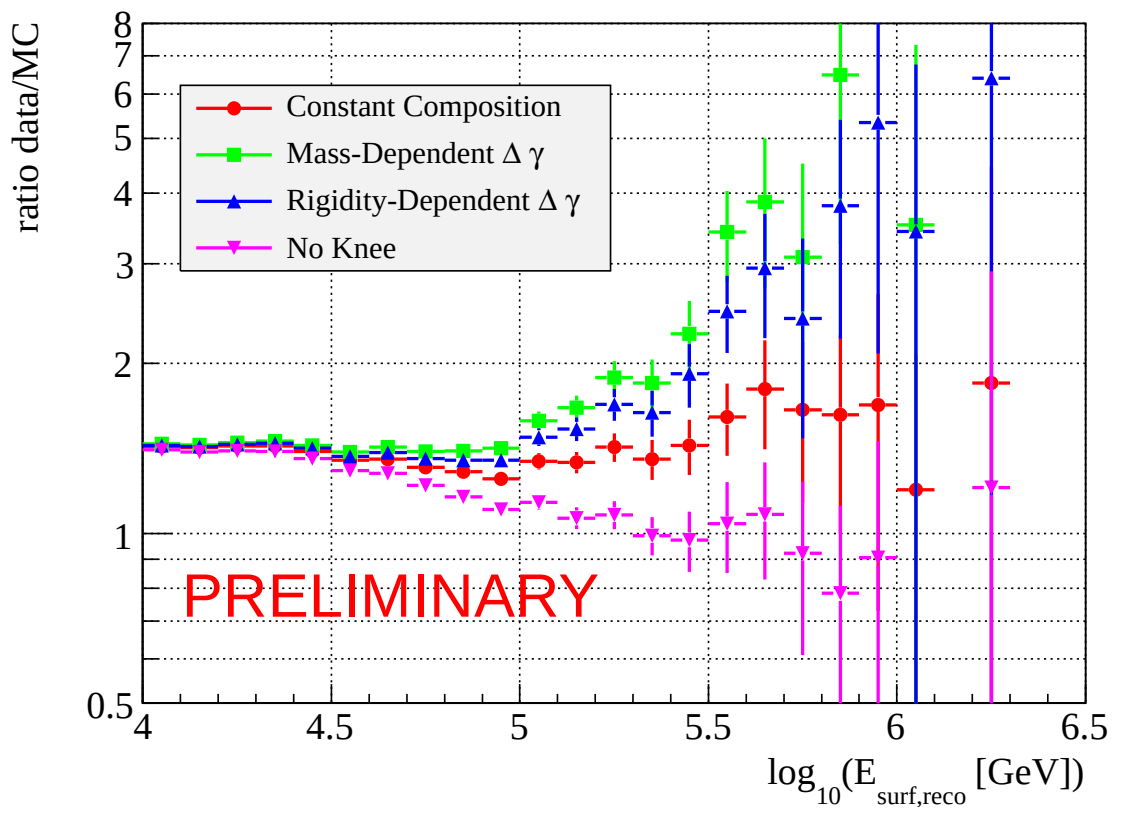

Figure 10: Demonstration of the composition sensitivity of the in-ice muon spectrum by measurements of the muon spectrum using the full year of IC59 data. Plotted is the ratio of data over the simulation prediction (details in [35]) as a function of the reconstructed muon surface energy, which is about a factor 10 lower than the primary cosmic ray energy. The comparison indicates a preference for either no change in the spectral slope around the knee ("no knee", not favoured by other observations) or the same slope change for all elements.

### 7.1 Cosmic Ray Physics with Muons in IceCube

**Atmospheric muon spectra:** Atmospheric muon and neutrino spectra measured with IceCube probe shower development of cosmic rays with primary energies above about 10 TeV. In a contribution to the conference [35] it was shown that with an accurate measurement of the muon spectra one can discriminate between different composition models (Fig. 10). At the current stage of the investigation a smoother transition of the different element contributions in the knee region (than for example suggested by the polygonato model [36]) is preferred. With additional systematic studies a clarification should be reached about what energy dependence of composition has to be used in simulation models.

This is a completely new approach to analyse cosmic ray composition in the knee region which is otherwise difficult to tackle. For the analysis new methods had to be developed, for example a method for determination of the energy of the leading muon by exploiting cascade-like stochastic energy losses [35].

**Laterally separated muons:** At high energies the muons reach the in-ice detector in bundles which are, for primaries above about 1 PeV, collimated within radii of the order of some 10 m. Most of the muons stem from the soft peripheral collisons with little transverse momentum transfer. Perturbative QCD calculations, however, predict the occurrance of muons with higher transverse momenta in some fraction of the events. A first analysis of the IC22/IT26 data [37], where the muon bundle was measured together with the shower energy in IceTop, demonstrated that separations of single muons from the bundle by more than about 100 m, corresponding to transverse momenta above about 6 GeV, could be detected. A better understanding of the remaining background from uncorrelated multiple events and an unfolding from the lateral separation to transverse momentum distributions is currently pursued. With a larger detector and also without requiring IceTop coverage, the statistics will be sufficient do perform a detailed analysis and comparison to the model predictions for meson production. This will have important implications for air shower simulations which the cosmic ray analyses have to rely on.

**Seasonal variations of the muon rate:** IceCube observes a $\pm 8\%$ seasonal variation of muon rates in the ice. This modulation is strongly correlated with the variability of the temperature, and thus of the density, in the upper atmosphere at heights corresponding to pressures around 10 to 100 hPa. The convolution of the density profile with the production cross section for muons defines the effective temperature $T_{\text{eff}}$. The relation between the effective temperature change and the rate change, assumed to be linear,

$$\frac{\Delta R_\mu}{\langle R_\mu \rangle} = \alpha_T \frac{\Delta T_{\text{eff}}}{\langle T_{\text{eff}} \rangle}, \qquad (3)$$

depends on the $K/\pi$ production ratio. From the coefficient $\alpha_T$ measured over 4 years on a sample of 150 billion events a preliminary result is reported in [38] which indicates that the currently assumed $K/\pi = 0.15$ has to be lowered to about 0.1. If confirmed, this would lead to modifications of the models for air shower simulation.

### 7.2 Atmospheric Neutrino Spectra

**Muon neutrinos:** IceCube has the most precise determination of the atmospheric muon neutrino spectrum at high energies (Fig. 7). This spectrum has to be unfolded from the measured muon energies to the neutrino energies. A measurement of electromagnetic and hadronic cascades generated by neutrinos of all flavours would yield a better energy determination. This is important especially at the high energy end where signals from diffuse neutrino fluxes are searched for. The potential transition from atmospheric to cosmic neutrinos is still theoretically uncertain due to missing information about composition and about the uncertainty in the prompt contribution from heavy quark production.

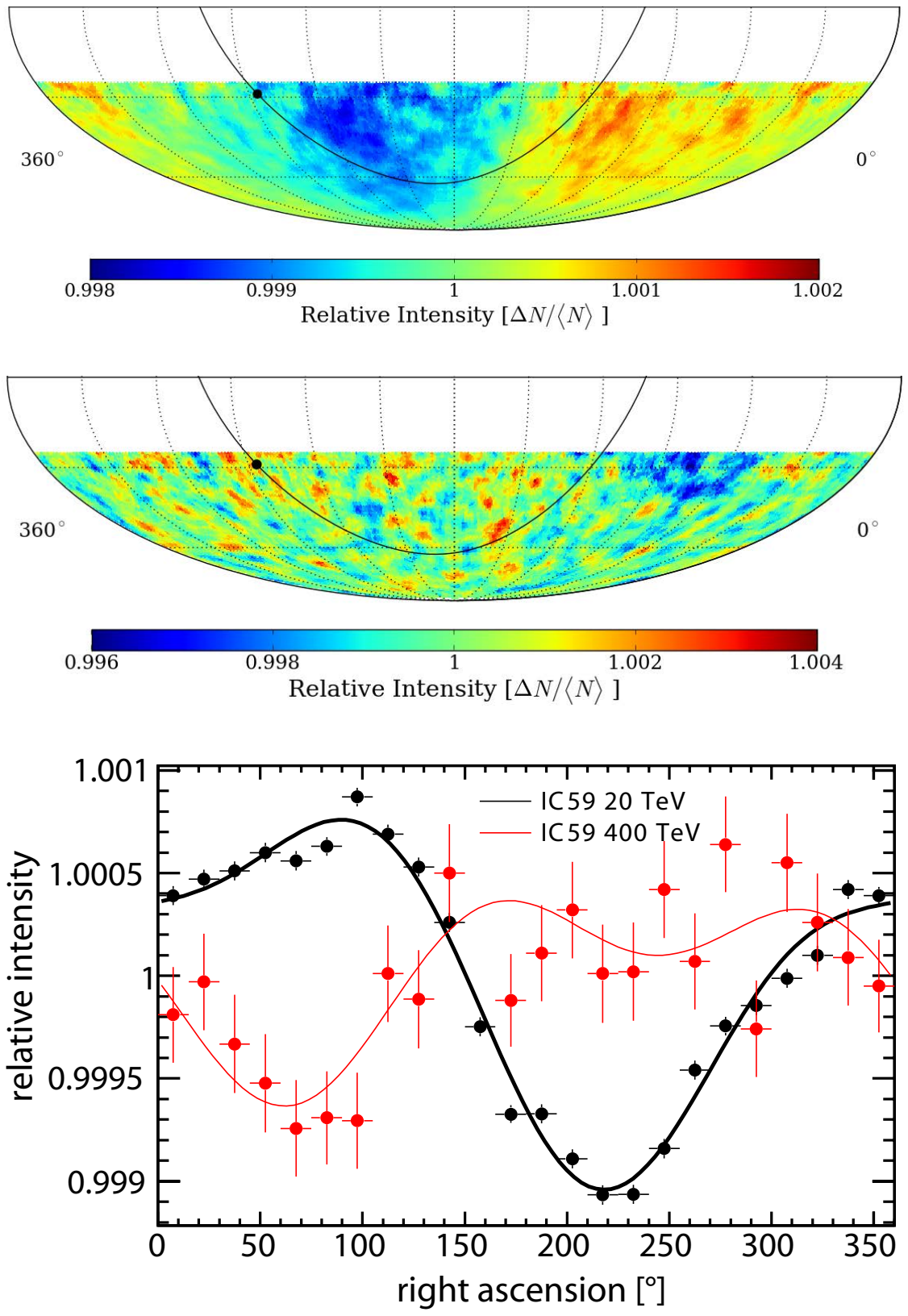

Figure 11: Relative intensity map in equatorial coordinates for cosmic rays of the 20-TeV sample (top) and the 400-TeV sample (middle). The curve indicates the galactic plane with the dot as the galactic center. The projections unto the right ascension of both maps are shown at the bottom.

**Cascade analysis with DeepCore:** In a first analysis of data taken with the DeepCore detector in the IC79 configuration (2010/11) cascades from atmospheric neutrinos have been analysed. In the DeepCore detector 1029 cascade candidates have been observed with a medium energy around 180 GeV for 281 days of data [10] while 1104 were predicted from simulations using the Bartol model [39]. Of the predicted events 59% are cascades with about equal amounts of $\nu_e$ CC and $\nu_\mu$ NC events. The remaining 41% is background from muon tracks from up-going $\nu_\mu$. Background from down-going atmospheric neutrinos and other systematic uncertainties are still under investigation [10].

This result from the newly commissionned DeepCore detector supports the expectations for the performance of the detector (Section 3). The physics goals of measuring neutrino oscillations [9], decreasing the mass range for the WIMP search and enhancing the sensitivity for supernovae detection become very realistic.

## 7.3 Cosmic Ray Anisotropy

IceCube has collected a large amount of cosmic ray muon events, about $10^{11}$ events between 2007 and 2010, and every year of running with the full detector will increase this number by about the same amount. These events have been used to study cosmic ray anisotropies, for the first time in the Southern sky. The observation of anisotropies on multiple angular scales has been previously reported [40, 41]. At this conference, analyses of anisotropies using $33 \cdot 10^9$ events from IC59 data were presented with results on energy and angular scale dependencies as well as various stability tests of the analyses [42, 43, 44].

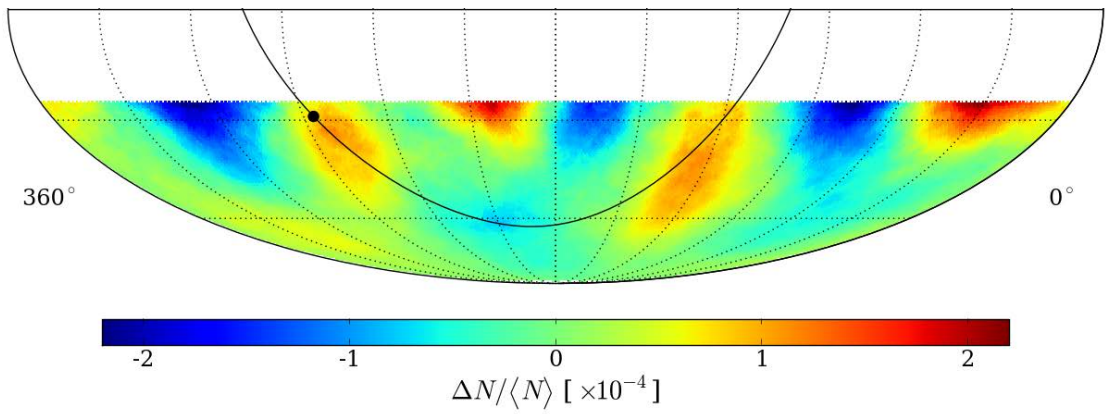

Figure 12: Cosmic ray anisotropies on the scale of 10 to 30°. Representation as for the maps in Fig. 11

Figure 11 shows skymaps of relative intensities for selections of muon energies resulting in primary energy distributions which center around 20 TeV and 400 TeV. In the 20-TeV right ascension projection a clear structure dominated by a dipole and quadrupole contribution is visible while the most significant feature in the 400-TeV data set is a deep deficite with a completely different phase than the dip in the 20-TeV data. For more details see [42].

In addition to large-scale features in the form of strong dipole and quadrupole moments, the data include several localized regions of excess and deficit on scales between 10° and 30° (Fig. 12). Angular decomposition into sperical harmonics exhibits significant contributions up to l=15. More details can be found in [43].

As yet the anisotropies observed on multiple angular scales and at different energies have not found an explanation. One could expect an effect due to the movement of the solar system relative to the Milky Way, the so-called Compton-Getting effect. This effect which results in a dipole component in the cosmic ray intensity distribution cannot, at least not fully, explain the data. Theoretical explanations like local magnet fields affecting the cosmic ray streams and/or nearby sources of cosmic rays are discussed. The determination of the energy dependence of anisotropies will be crucial for scrutinizing models. For this reason an analysis using IceTop with a better energy resolution and an extension to the PeV range for the primary cosmic rays has been started.

## 7.4 Cosmic Ray Composition

As mentioned above, the combination of the in-ice detector with the surface detector offers a unique possibility to determine the spectrum and mass composition of cosmic rays from about 300 TeV to 1 EeV.

The first analysis exploiting the IceTop-IceCube correlation was done on a small data set corresponding to only one month of data taken with about a quarter of the final detector [45]. The energy was restricted to 1 to 30 PeV. A neural network was employed to determine from the measured input variables shower size and muon energy loss the primary energy and mass (Fig. 13). The resulting average logarithmic mass is shown in Fig. 14. These results are still dominated by systematic uncertainties, such as the energy scale of the muons in IceCube and of the effects of snow accumulation on the IceTop tanks.

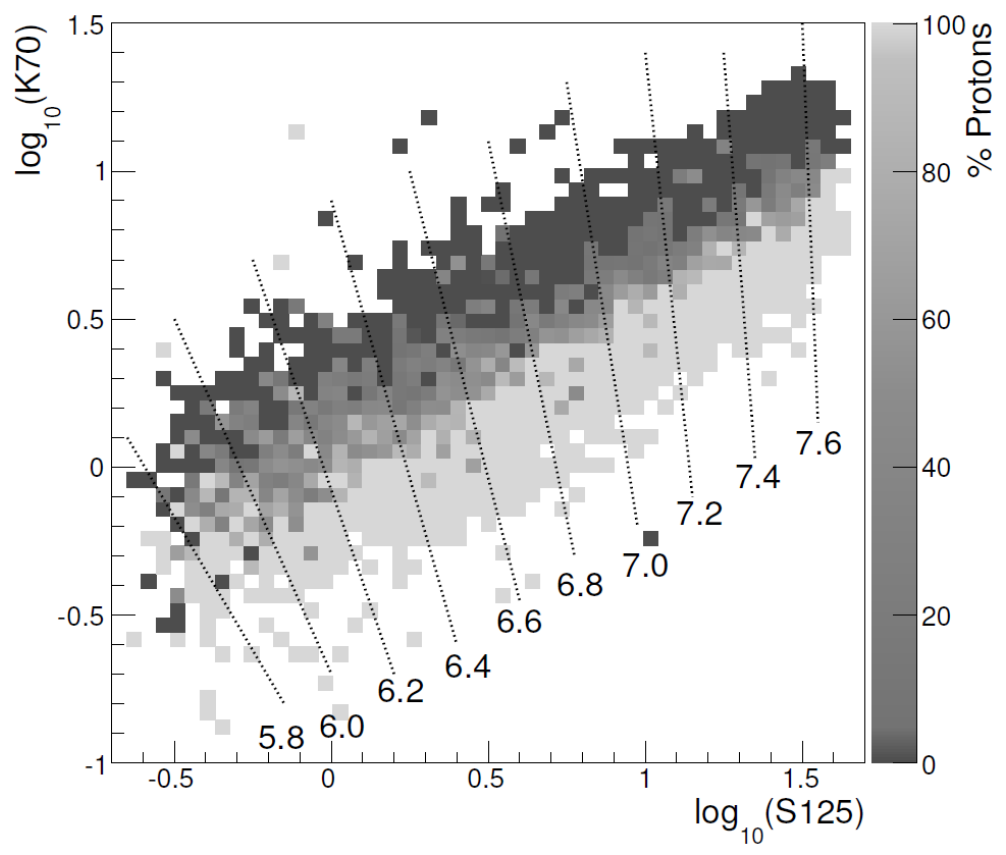

Figure 13: Simulated correlation between the energy loss of the muon bundels in the ice (K70) and the shower size at the surface (S125) for proton and iron showers. The shading indicates the percentage of protons over the sum of protons and iron in a bin. The lines of constant primary energy are labelled with the logarithms of the energies.

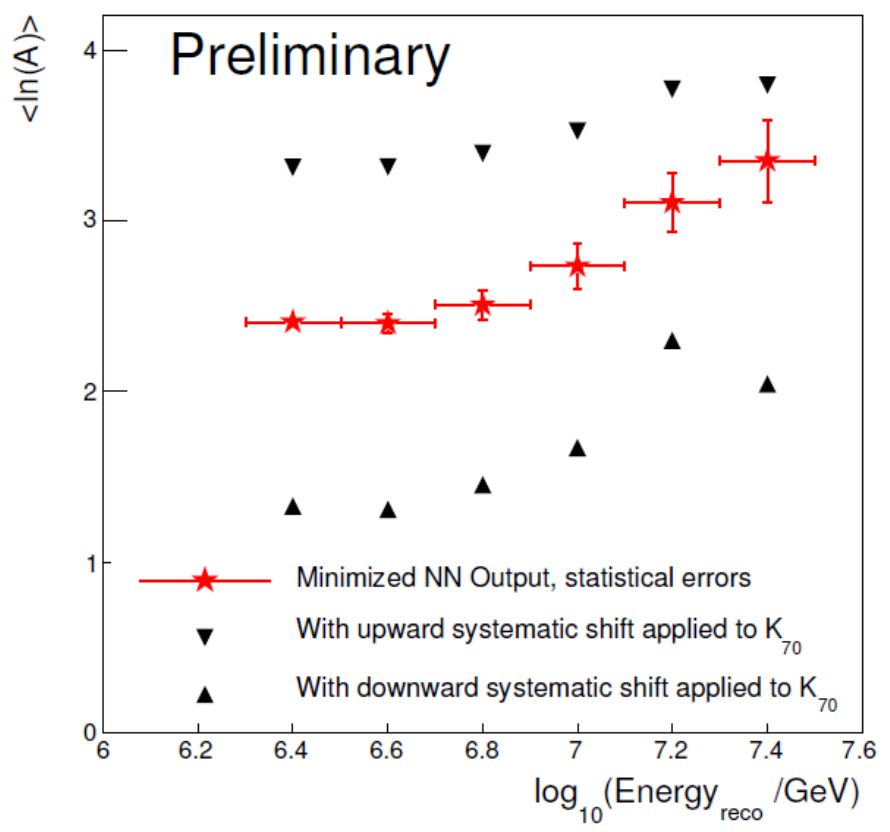

Figure 14: Average logarithmic mass of primary cosmic rays measured with IC40/IT40.

A first look into the IC79/IT73 data set taken in 2010 shows that there will be enough statistics for composition analysis up to at least 1 EeV [46]. An estimation yields about 150 events with energies larger than 300 PeV and 15 events larger than 1 EeV in 1 year of data taking with the full detector.

The systematic uncertainties related to the models can be reduced by including different mass sensitive variables, like zenith angle dependence of shower size [11], muon rates in the surface detector and shower shape variables, and checking for consistency.

The IceCube-IceTop combination has also been used to identify high-energy photons as IceTop showers with no muons in the ice [47]

## 8 Transient Rate Monitoring

Transient events such as supernovae, GRBs or sun flares, if they generate very high fluxes of low energy particles, could be observed as general rate increases above the noise level in the DOMs even if they could not be detected individually by IceCube or IceTop.

Supernova explosions in our and nearby galaxies would be observable by IceCube via a rate increase in all DOMs due to a high interaction rate of low energy neutrinos. With a rather low average noise of 286 Hz per DOM (with afterpulse rejection) IceCube is particularly suited to emit supernova alerts, specifically important when the supernova is obscured by dust or stars in a dense region. Measurements would be sensitive to the supernova parameters such as the progenitor star mass, neutrino oscillations and hierarchy. In the contribution [48] possibilities for improving the current sensitivities, including also DeepCore, are discussed.

IceTop is able to monitor cosmic ray products from transient events such as from Sun flares, as demonstrated with the observation of the Dec 13, 2006 Sun flare event [49]. The detector readout has since then been setup such that counting rates could be obtained at different thresholds allowing to unfold cosmic ray spectra during a flare. At this conference the observation of a Forbush decrease in February 2011 was reported [50].

## 9 Summary and Outlook

The IceCube Neutrino Observatory has been completed and reached or exceeded its design sensitivity. As yet, results from the partly completed detector (IC22,40,59) show no evidence for cosmic neutrinos, although the detector reached sensitivities which are either close to model predictions or are sometimes seriously challenging models. Most notable is the IC40+59 limit on GRBs which is 5 times below the model prediction of [16] with preferred parameters, demanding a reassessment of the model and/or parameters. Point source searches, time dependent or not, with and without candidate lists, have not yet reached the level to constrain the most common models, but will in some years of running. The hope is to accelerate the progress by further developing methods to enhance significances, for example by employing multi-messenger methods and follow-up programs with optical, X-ray and $\gamma$-ray telescopes.

The limits on diffuse cosmic neutrino fluxes are now a factor of 2.5 below the Waxman-Bahcall bound, indicating that the limits have reached a relevant region of predictions. The reported analyses of cascade events promise the opening of a new window for studies of atmospheric neutrinos, in particular their 'prompt' contributions, and of cosmic neutrinos with good energy resolution. In the EHE region the sensitivity to the range of GZK predictions will be reached within a few years.

Concerning searches for 'exotic' particles, limits for WIMP masses between 50 GeV and 5 TeV have reached regions in the parameter space which are not excluded by direct search experiments. Magnetic monopole limits are now nearly a factor 1000 below the 'Parker Bound' (upper bound derived from the strength of existing cosmic magnetic fields) and are constraining GUT models.

Although most of these limits are very important and unique complements to results with other messengers it is comforting to know that also positive observations have been made with IceCube. These results concern mainly the field of cosmic rays and are mostly of high importance for the improvement of cosmic ray and airshower models. Results have been reported on atmospheric neutrino and muon spectra, muons with large transverse momenta, cosmic ray composition and cosmic ray anisotropies on multiple angular scales. The cosmic ray anisotropies, the first time measured in the Southern sky, are drawing a lot of interest but have not yet found an explanation.

IceCube can be used as a unique instrument to measure transient events, such as supernovae, GRBs and sun flares. This already led to results on heliosperic physics.

Looking into the future: it seems as if the discovery of cosmic high energy neutrinos might need some more years, in which the existing detectors will be exploited, improved and extended. The first, already accomplished, extension was DeepCore opening the way to low energy phenomena such as neutrino oscillations, low mass WIMPs and supernova physics. A new low energy extension with very dense optical sensor instrumentation to allow for Cherenkov imaging in a megaton scale detector is studied, an interesting physics application being the search for proton decay [51]. At the high energy end: radio and acoustic extensions are studied to reach the sensitivity for GZK neutrino fluxes [52, 53] and to extend the air shower detection capabilities [54].

**Acknowledgements:** I would like to thank the many people in the IceCube Collaboration who helped me preparing the talk and the proceedings.

## References

[1] The IceCube Collaboration, The IceCube Neutrino Observatory I-VI: IceCube ICRC 2011 contributions:
I: Point Source Searches, arXiv:1111.2741; II: All Sky Searches: Atmospheric, Diffuse and EHE, arXiv:1111.2736; III: Cosmic Rays, arXiv:1111.2735; IV: Searches for Dark Matter and Exotic Particles, arXiv:1111.2738; V: Future Developments, arXiv:1111.2742; VI: Neutrino Oscillations, Supernova Searches, Ice Properties, arXiv:1111.2731.

[2] Felix Aharonian, "Probing cosmic ray accelerators with gamma rays and neutrinos", invited talk, these proceedings.

[3] IceCube Collab., ICRC contr. 807, arXiv:1111.2735.

[4] R. Abbasi et al. (IceCube), NIM A601 (2009) 294.

[5] IceCube Collab., ICRC contr. 333, arXiv:1111.2731.

[6] IceCube Collab., ICRC contr. 899, arXiv:1111.2735.

[7] IceCube Collab., ICRC contr. 1235, arXiv:1111.2741.

[8] R. Abbasi et al. (IceCube), PRD 84 (2011) 082001.

[9] IceCube Collab., ICRC contr. 329, arXiv:1111.2731.

[10] IceCube Collab., ICRC contr. 324, arXiv:1111.2736.

[11] F. Kislat, Astrophys. Space Sci. Trans. 7 (2011) 175.

[12] R. Abbasi, et al. (IceCube), ApJ 732 (2011) 18.

[13] IceCube Collab., ICRC contr. 784, arXiv:1111.2741.

[14] R. Abbasi, et al. (IceCube), arXiv:1104.0075.

[15] E. Waxman. NP B Proc. Suppl., 118 (2003) 353.

[16] D. Guetta et al., ApP 20 (2004) 429.

[17] IceCube Collab., ICRC contr. 764, arXiv:1111.2741.

[18] R. Abbasi et al. (IceCube), PRL 106 (2011) 141101.

[19] IceCube Collab., ICRC contr. 445, arXiv:1111.2741.

[20] IceCube Collab., ICRC contr. 535, arXiv:1111.2741.

[21] IceCube Collab., ICRC contr. 334, arXiv:1111.2741.

[22] IceCube Collab., ICRC contr. 736, arXiv:1111.2736.

[23] Ahlers et al., ApP 34 (2010) 106.

[24] E. Waxman and J. Bahcall, PRD 59 (1998) 023002.

[25] IceCube Collab., ICRC contr. 759, arXiv:1111.2736.

[26] IceCube Collab., ICRC contr. 1097, arXiv:1111.2736.

[27] IceCube Collab., ICRC contr. 949, arXiv:1111.2736.

[28] R. Abbasi, et al. (IceCube), PRD 83 (2011) 092003.

[29] IceCube Collab., ICRC contr. 773, arXiv:1111.2736.

[30] IceCube Collab., ICRC contr. 778, arXiv:1111.2736.

[31] IceCube Collab., ICRC contr. 327, arXiv:1111.2738.

[32] IceCube Collab., ICRC contr. 1178, arXiv:1111.2738.

[33] IceCube Collab., ICRC contr. 1024, arXiv:1111.2738.

[34] IceCube Collab., ICRC contr. 734, arXiv:1111.2738.

[35] IceCube Collab., ICRC contr. 85, arXiv:1111.2735.

[36] J. R. Hörandel, ApP 19 (2003) 193.

[37] IceCube Collab., ICRC contr. 323, arXiv:1111.2735.

[38] IceCube Collab., ICRC contr. 662, arXiv:1111.2735.

[39] G. D. Barr et al., Phys. Rev. D70 (2004).

[40] R. Abbasi et al. (IceCube), ApJ 718 (2010) L194.

[41] R. Abbasi et al. (IceCube), ApJ 740 (2011)16.

[42] IceCube Collab., ICRC contr. 305, arXiv:1111.2735; Abbasi et al. (IceCube) arXiv:1109.1017, acc. by ApJ.

[43] IceCube Collab., ICRC contr. 306, arXiv:1111.2735.

[44] IceCube Collab.,ICRC contr. 308, arXiv:1111.2735.

[45] IceCube Collab., ICRC contr. 923, arXiv:1111.2735.

[46] IceCube Collab., ICRC contr. 838, arXiv:1111.2735.

[47] IceCube Collab., ICRC contr. 939, arXiv:1111.2735.

[48] IceCube Collab., ICRC contr. 1137, arXiv:1111.2731.

[49] R. Abbasi et al. (IceCube), ApJ Lett. 689 (2008) 65.

[50] IceCube Collab., ICRC contr. 921, arXiv:1111.2735.

[51] IceCube Collab., ICRC contr. 325, arXiv:1111.2742.

[52] IceCube Collab., ICRC contr. 1236, arXiv:1111.2742.

[53] IceCube Collab., ICRC contr. 316, arXiv:1111.2742.

[54] IceCube Collab., ICRC contr. 1102, arXiv:1111.2742.